\documentclass[%
a4paper,
preprint,
aps,prb,%
superscriptaddress,%
showpacs,showkeys]{revtex4-1}

\usepackage[english]{babel}
\usepackage[intlimits]{amsmath}
\usepackage{amssymb,graphicx}
\usepackage{ifpdf}
\ifpdf\usepackage{epstopdf}\fi

\begin{document}

\title{Photoluminescence of a quantum-dot molecule}

\author{Stanislav~Yu.~Kruchinin}
\email{stanislav.kruchinin@mpq.mpg.de}
\affiliation{Max Planck Institute of Quantum Optics, Hans-Kopfermann-Stra\ss{}e 1, D-85748 Garching, Germany}

\author{Ivan~D.~Rukhlenko}
\affiliation{Advanced Computing and Simulation Laboratory (A$\chi$L), Department of Electrical and Computer Systems Engineering, Monash University 3800, Victoria, Australia}

\author{Anvar~S.~Baimuratov}
\affiliation{Saint Petersburg National Research University of Information Technologies, Mechanics and Optics, 197101 St. Petersburg, Russia}

\author{Mikhail~Yu.~Leonov}
\affiliation{Saint Petersburg National Research University of Information Technologies, Mechanics and Optics, 197101 St. Petersburg, Russia}

\author{Vadim~K.~Turkov}
\affiliation{Saint Petersburg National Research University of Information Technologies, Mechanics and Optics, 197101 St. Petersburg, Russia}

\author{Yurii~K.~Gun'ko}
\affiliation{Saint Petersburg National Research University of Information Technologies, Mechanics and Optics, 197101 St. Petersburg, Russia}
\affiliation{School of Chemistry and CRANN Institute, Trinity College, Dublin, Dublin 2, Ireland}

\author{Alexander~V.~Baranov}
\affiliation{Saint Petersburg National Research University of Information Technologies, Mechanics and Optics, 197101 St. Petersburg, Russia}

\author{Anatoly~V.~Fedorov}
\affiliation{Saint Petersburg National Research University of Information Technologies, Mechanics and Optics, 197101 St. Petersburg, Russia}


\begin{abstract}
We formulate a theory of low-temperature, stationary photoluminescence from a quantum-dot molecule composed of two spherical quantum dots whose electronic subsystems are resonantly coupled \emph{via} the Coulomb interaction.
We show that the coupling leads to the hybridization of the first excited states of the quantum dots, manifesting itself as a pair of photoluminescence peaks with intensities and spectral positions strongly dependent on the geometric, material, and relaxation parameters of the quantum-dot molecule.
These parameters are explicitly contained in the analytical expression for the photoluminescence differential cross section derived in the paper.
The developed theory and expression obtained are essential in interpreting and analyzing spectroscopic data on the secondary emission of coherently coupled quantum systems.
\end{abstract}


\maketitle

\section{Introduction}

The nonradiative transfer of energy in low-dimensional structures has been the subject of much research~\cite{Brichkin_2013_HighEnergyChem_47_277,Rogach_2011_NanoToday_6_355} due to the many prospective uses of this phenomenon in optoelectronics~\cite{Guzelturk_2013_LPR_8_73,Demir_2011_NanoToday_6_632},
quantum computing~\cite{Basset_2013_PRB_88_125312,Taylor_2005_NaturePhys_1_177},
biology and medicine~\cite{Zrazhevskiy_2010_CSR_39_4326,Beljonne_2009_JPCB_113_6583}.
The first theory of nonradiative energy transfer was developed in the late 1940s by F\"orster~\cite{Forster_1948_AP_2_55}, who studied the resonant migration of energy between a pair of dye molecules using the semiclassical quantum approach while considering the dipole--dipole interaction between the molecules.
F\"orster's theory was extended a few years later by Dexter~\cite{Dexter_1953_JCP_21_836} to include transfer by means of dipole-forbidden transitions, which occur due to the overlapping of the dipole field of a sensitizer with the quadrupole field of an activator and exchange effects.
The past decade has witnessed the emergence of many experimental\cite{Sarkar_2013_JPCC_117_21988,Lin_2012_ACSNANO_6_4029,Lunz_2011_PRB_83_115423,
Lunz_2010_PRB_81_205316} and theoretical\cite{Belyakov_2013_PRB_88_045439,Kruchinin_2010_PRB_81_245303,
Kruchinin_2010_JCP_133_104704,Kruchinin_2008_PRB_78_125311,Allan_2007_PRB_75_195311,
Rukhlenko_2006_OS_100_238,Rukhlenko_2006_OS_101_253} works devoted to the investigation of various aspects of nonradiative energy exchange between semiconductor quantum dots (QDs), including the studies on how the exchange is affected by the nearby metallic nanoparticles~\cite{West_2012_JPCC_116_20496,Rukhlenko_2009_OE_17_17570} and photon modes of optical microcavities~\cite{Minkov_2013_PRB_87_125306,Majumdar_2012_PRB_85_195301}.
The widespread interest in different kinds of QD nanostructures --- including molecules and oligomers~\cite{Daniels_2013_PRB_88_205307,Pont_2013_PRB_88_241304}, two- and three-dimensional supercrystals~\cite{Baimuratov_2013_SciRep_3_1727,Baimuratov_2013_OL_38_2259,Rogach_2002_AFM_12_653,Murray_2000_ARMS_30_545}, as well as dendrites~\cite{Sukhanova_2007_Nanotechnology_18_185602} --- is explained by the size-dependent energy spectrum of QDs, their high chemical stability and fluorescence brightness (the product of the quantum yield and extinction coefficient).
These features make QD nanostructures ideal objects for experimental studies of the nonradiative energy transfer \emph{via} the methods of optical spectroscopy.
Of significance from the theoretical viewpoint is that in many practical instances such transfer can be adequately described within the framework of the dipole--dipole approximation even when the QDs almost touch each other~\cite{Kruchinin_2008_PRB_78_125311,Baer_2008_JCP_128_184710,Curutchet_2008_JPCC_112_13336,Allan_2007_PRB_75_195311}.

The interdot Coulomb interaction can lead to both the incoherent and coherent energy transfers in the closely packed assemblies of QDs, just as it does in atomic and molecular systems~\cite{Agranovich_1982}.
The presence or absence of coherence effects in a QD dimer is determined by the relationship between the interdot-interaction matrix element $M_{\mathrm{I,II}}$ (subscripts I and II correspond to the first and second QDs), energy detuning $\Delta_\mathrm{I,II}=E_\mathrm{I}-E_\mathrm{II}$ of the QDs' excitations coupled by the interaction, and dephasing rate $\Gamma$ of the interdot transitions.
The formation of the entangled states of the dimer and the coherent energy transfer between the QDs are possible when
\begin{equation}\label{e:GDOmega}
  |M_{\mathrm{I,II}}| \gg |\Delta_\mathrm{I,II}|, \hbar\Gamma.
\end{equation}
Otherwise, only the incoherent energy transfer, either reversible or not, can occur~\cite{Kruchinin_2010_JCP_133_104704,Kruchinin_2010_PRB_81_245303,Kruchinin_2008_PRB_78_125311}.

The value of $\Gamma$ can be approximated by a sum of the dephasing rates of electronic transitions in the first and second QDs,
\begin{gather*}
  \Gamma\approx\Gamma_\mathrm{I}+\Gamma_\mathrm{II},\quad
  \Gamma_\alpha = (\gamma_{i,\alpha} + \gamma_{f,\alpha})/2
  +\bar\gamma_{fi,\alpha}\quad(\alpha = \mathrm{I},\mathrm{II}),
\end{gather*}
where $\gamma_{i,\alpha}$ and $\gamma_{f,\alpha}$ are the energy relaxation rates of the initial and final transition states and $\bar\gamma_{if,\alpha}$ is the pure dephasing rate of transition $|i,\alpha\rangle\to|f,\alpha\rangle$.
The matrix elements of the dipole-allowed transitions are typically less than a few millielectronvolts in QDs made of direct-bandgap semiconductors~\cite{Kruchinin_2010_JCP_133_104704,Curutchet_2008_JPCC_112_13336,Allan_2007_PRB_75_195311}, which means that the coherent coupling can only be realized between QDs with relatively slow phase and energy relaxations.
A simple estimate of the dephasing rate $\Gamma$ shows that the condition in Eq.~\eqref{e:GDOmega} is satisfied for a pair of identical CdSe QDs at temperatures below 90~K.
Since the matrix elements of the dipole-forbidden transitions are much smaller than those of the dipole-allowed one, the forbidden interdot interaction is far less attractive from the viewpoint of experimental investigation of the coherence effects.

Another problem associated with the realization of the coherent energy transfer in QD systems is the variation of the intraband energy relaxation rates in QDs over a wide range of $10^9$ to $10^{13}$~s$^{-1}$ even at cryogenic temperatures~\cite{Pandey_2008_SCI_332_929,Bonati_2007_PRB_76_033304,Hendry_2006_PRL_96_057408,
Guyot-Sioneest_2005_JCP_123_074709,Fedorov_2003_PRB_68_205318,Fedorov_2003_SSC_128_219}.
Therefore, even if the resonance condition is satisfied for the fundamental transition of one QD and some transition between the high-energy excited states of the other, the coherent coupling may still be absent due to the fast intraband relaxation.
We can thus conclude that the coherent coupling should be the easiest to achieve between the lowest-energy states of the QDs made of the wide-bandgap semiconductors, because the interband relaxation rates of the fundamental transitions in such QDs can be much less than $|M_{\mathrm{I,II}}|/\hbar$.~\cite{Kruchinin_2008_PRB_78_125311,Kruchinin_2010_JCP_133_104704,Kruchinin_2010_PRB_81_245303}

One of the key tasks in the field of nonradiative energy transfer is the development of a theoretical framework of the photoluminescence spectroscopy that would enable distinguishing between different regimes of energy transfer in QD nanostructures and extracting important QD parameters (e.g. energy spectrum and phase relaxation rates) from experimental data.
In our previous works~\cite{Kruchinin_2010_PRB_81_245303,Kruchinin_2008_PRB_78_125311}, we theoretically studied stationary photoluminescence from the double QDs exhibiting the reversible or nonreversible incoherent resonant energy transfer.
This paper continues these studies by presenting a theory on the secondary emission from a pair of coherently coupled QDs and analyzing the manifestations of coherence effects in the photoluminescence spectra.

\section{Hamiltonian formalism}

Consider a QDM whose interaction with the classical excitation field and the quantum radiation field is described by the Hamiltonian
\begin{equation}\label{e:H}
  H = H_\mathrm{QDM}+H_\mathrm{R}+H_\mathrm{QDM,L}+H_\mathrm{QDM,R},
\end{equation}
where the first two terms represent the noninteracting QDM and emitted photons whereas the rest describes the interactions.

We focus on the QDM composed of two QDs and described by the Hamiltonian
\begin{equation}\label{e:HQDM}
  H_\mathrm{QDM}=\sum_\alpha\sum_{p}E_{p,\alpha}^{\vphantom{\dagger}}
  a_{p,\alpha}^\dagger a_{p,\alpha}^{\vphantom{\dagger}} +\sum_{p,\,q}\left(M_{q\mathrm{I},p\mathrm{II}}^{\vphantom{\dagger}}
  a_{q,\mathrm{I}}^+a_{p,\mathrm{II}}^{\vphantom{\dagger}}+\mathrm{H.c.}\right),
\end{equation}
where $E_{p,\alpha}$ is the energy of the electron--hole-pair state $p$ in the first $(\alpha = \mathrm{I})$ or second $(\alpha=\mathrm{II})$ QD, and $a_{p,\alpha}^\dagger$ and $a_{p,\alpha}^{\vphantom{\dagger}}$ are the creation and annihilation operators of the electron--hole pairs.
The matrix element $M_{q\mathrm{I},p\mathrm{II}}\equiv\langle q,\mathrm{I}|V_\mathrm{C}|p,\mathrm{II}\rangle$ describes the Coulomb interaction between the QDs, which are coupled through the screened potential
\begin{equation}\label{e:colp}
    V_\mathrm{C}(\mathbf{r}, \mathbf{r}_{\mathrm{I}}, \mathbf{r}_{\mathrm{II}})
    =\frac{e^2}{\varepsilon|\mathbf{r} + \mathbf{r}_{\mathrm{I}} - \mathbf{r}_{\mathrm{II}}|},
\end{equation}
where $\mathbf{r}$ is the vector directed from the center of the second QD to the center of the first QD whereas ${\bf r}_\mathrm{I}$ and ${\bf r}_\mathrm{II}$ are the radius vectors of electrons in the reference frames with the origins at the QD centers.
By considering spherical QDs in a dielectric matrix, one can describe the effect of screening by the effective permittivity~\cite{Allan_2007_PRB_75_195311,Kruchinin_2010_JCP_133_104704}
\begin{equation*}
    \varepsilon=\frac{(\varepsilon_{\mathrm{I}}+ 2\varepsilon_\mathrm{M})(\varepsilon_{\mathrm{II}}+
    2\varepsilon_\mathrm{M})}{9\varepsilon_\mathrm{M}},
\end{equation*}
where $\varepsilon_\mathrm{I}$, $\varepsilon_\mathrm{II}$, and $\varepsilon_\mathrm{M}$ are the high-frequency permittivities of the QDs and matrix.
Note that Eq.~\eqref{e:HQDM} neglects the interdot exchange interaction due to its weakness for QDMs embedded in dielectric~\cite{Franceschetti_1997_PRL_78_915}.

The Hamiltonian of noninteracting photons is of the form
\begin{equation*}
    H_{\mathrm{R}}=\sum_k\hbar\omega_k^{\vphantom{+}}
    b_k^+b_k^{\vphantom{+}},
\end{equation*}
where $b_k^+$ and $b_k^{\vphantom{+}}$ are the creation and annihilation operators of photons of mode $k$ and frequency $\omega_k$, whereas the last two terms in Eq.~\eqref{e:H} are given by
\begin{equation*}
  H_{\mathrm{QDM,R}}
  =\sum_\alpha\sum_{p,\,k}g_{\alpha,k}\left(i\hbar V_{p\alpha,0\alpha}^{(k)}
  b_k^{\vphantom{+}}a_{p,\alpha}^++\mathrm{H.c.}\right)
\end{equation*}
and
\begin{equation*}
  H_{\mathrm{QDM,L}}=\sum_\alpha\sum_p\left(\phi(t)
  V_{p\alpha,0\alpha}^{(\mathrm{L})}e^{-i\omega_\mathrm{L}t}
  a_{p,\alpha}^++\mathrm{H.c.}\right),
\end{equation*}
where $g_{\alpha,k}=\sqrt{2\pi\hbar\omega_k/(\varepsilon_\alpha V)}$, $V$ is the normalization volume, $V_{p\alpha,0\alpha}^{(\eta)}=-\,e\langle p,\alpha|\mathbf r\mathbf e_\eta|0,\alpha\rangle$ $(\eta=\mathrm{L},k)$, $-e\mathbf r$ is the dipole moment operator, $\mathbf e_\eta$ is the polarization vector, and $\phi(t)$ is the complex envelope of the excitation field of frequency $\omega_\mathrm{L}$.

\section{Resonant coupling of quantum dots}

As was mentioned earlier, the coherent coupling of QDs in a QDM is strongest when the energies of the lowest excited electronic states of the QDs coincide.
If this resonance condition is nearly satisfied, then the interdot interaction is dominated by two resonant terms and the Hamiltonian of the QDM takes the form
\begin{gather*}
  H_\mathrm{QDM}=E_\mathrm{I}^{\vphantom{+}}
  a_\mathrm{I}^+a_\mathrm{I}^{\vphantom{+}}
  +E_\mathrm{II}^{\vphantom{+}}
  a_\mathrm{II}^+a_\mathrm{II}^{\vphantom{+}}
  +\langle01|V_\mathrm{C}|10\rangle a_\mathrm{II}^+a_\mathrm{I}^{\vphantom{+}}+
  \langle10|V_\mathrm{C}|01\rangle a_{\mathrm{I}}^+ a_{\mathrm{II}}^{\vphantom{+}},
\end{gather*}
where $a_\alpha^+$ and $a_\alpha^{\vphantom{+}}$ are the creation and annihilation operators of the electron--hole pairs in the lowest excited states of energies $E_\mathrm{I}$ and $E_\mathrm{II}$ and we have employed the following notations for the wave functions of the noninteracting QDs: $|00\rangle=|\mathrm{0,I}\rangle|\mathrm{0,II}\rangle$, $|10\rangle=|\mathrm{1,I}\rangle |\mathrm{0,II}\rangle$, and
$|01\rangle=|\mathrm{0,I}\rangle |\mathrm{1,II}\rangle$.

In order to describe the electronic subsystem of the QDM, we use the approximation of the infinitely high potential barriers for the confined electrons and holes, and the two-band model of the QD band structure~\cite{Kruchinin_2010_JCP_133_104704,Kruchinin_2010_PRB_81_245303,Kruchinin_2008_PRB_78_125311}.
We also assume that both QDs are in the regime of strong confinement and that their resonant interband transitions are dipole-allowed.
Then the matrix element of the Coulomb potential is found (in the dipole--dipole approximation) to be given by~\cite{Kruchinin_2010_PRB_81_245303}
\begin{equation}\label{e:M}
    M_{\mathrm{I,II}}\equiv\langle 10|V_\mathrm{C}|01\rangle=\frac{e^2\chi}{\varepsilon r^3}\big|\mathbf r_{vc}^{(\mathrm{I})}\big|\big|\mathbf r_{cv}^{(\mathrm{II})}\big|,
\end{equation}
where $\chi$ describes the orientational dependence of $M_{\mathrm{I,II}}$ and $\mathbf r_{cv}^{(\alpha)}$ is the matrix element of the coordinate operator.
By adopting the spherical coordinates with the $z$ axis parallel to vector $\mathbf r$, one gets the following functional dependency:\cite{Kruchinin_2010_PRB_81_245303}
\begin{equation}\label{e:chi}
    \chi(\theta_{\mathrm{I}},\theta_{\mathrm{II}},\varphi)
    =\sin\theta_{\mathrm{I}}\sin\theta_{\mathrm{II}} \cos\varphi-2\cos\theta_{\mathrm{I}} \cos\theta_{\mathrm{II}},
\end{equation}
where we have assumed that $\mathbf r_\mathrm{I}$ and $\mathbf r_\mathrm{II}$ make angles $\theta_\mathrm{I}$ and $\theta_\mathrm{II}$ with $\mathbf r$, and $\varphi$ is the difference between the azimuths of $\mathbf r_\mathrm{I}$ and $\mathbf r_\mathrm{II}$.

When the nearly resonant excitations of the two QDs are coupled through the Coulomb potential, they get hybridized and form excitations of the QDM.
The states of the new excitations are the superpositions of the QD states and can be found \emph{via} the canonical transformation technique~\cite{Davydov_1976}.
Annihilation of QDM excitations is described by new operators, $a_1$ and $a_2$, related to the old ones as
\begin{equation}\label{e:transf}
    \Bigg(
      \begin{array}{c}
        a_1 \\
        a_2
      \end{array}
    \Bigg)=
    \Bigg(
      \begin{array}{cc}
        \cos\vartheta & \sin\vartheta \\
        \cos\vartheta & -\sin\vartheta
      \end{array}
    \Bigg)
    \Bigg(
      \begin{array}{c}
        a_{\mathrm{I}} \\
        a_{\mathrm{II}}
      \end{array}
    \Bigg),
\end{equation}
where the transformation angle $\vartheta=(1/2)\arctan[2 M_{\mathrm{I,II}}/(E_{\mathrm{I}}-E_{\mathrm{II}})]$ $(-\pi/4<\vartheta<\pi/4)$ is the parameter of the perturbation theory for a pair of degenerate states.
The wave functions and energies of the QDM excitations are given by
\begin{gather*}
   |1\rangle\rangle=|10\rangle\cos\vartheta+|01\rangle\sin\vartheta,\\
   |2\rangle\rangle=|01\rangle\cos\vartheta-|10\rangle\sin\vartheta,
\end{gather*}
and
\begin{equation}\label{e:03}
    E_{1,2}\equiv\hbar\omega_{1,2}=\frac{1}{2}\left(E_{\mathrm{I}}
    +E_{\mathrm{II}}\pm\sqrt{(E_{\mathrm{I}}-E_{\mathrm{II}})^2
    +4|M_{\mathrm{I,II}}|^2}\right).
\end{equation}

\begin{figure}[!ht]
\includegraphics[width=7cm]{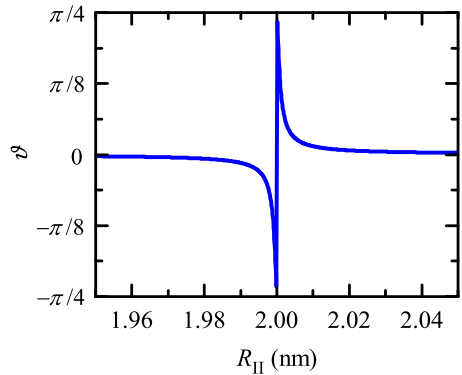}
\caption{%
(Color online) Transformation angle $\vartheta$ \emph{vs} radius $R_{\mathrm{II}}$ of the second QD for $R_{\mathrm{I}}=2$~nm.
}
\label{f:theta_RII}
\end{figure}

The energy splitting $E_1-E_2$ of states $|1\rangle\rangle$ and $|2\rangle\rangle$ is relatively small and very sensitive to the materials, shapes, and sizes of the QDs.
The latter can be seen from the behavior of parameter $\vartheta$ as a function of the second QD radius illustrated in Fig.~\ref{f:theta_RII}.
Since the radii of real QDs vary discretely, with the steps determined by the lattice constants of the QD materials, careful control of the radii, materials, and shapes of the QDM components is required in order to achieve the resonance and realize the coherent coupling even at cryogenic temperatures.

The interaction between the states of the QDM with coherently coupled QDs and the classical excitation field is described by the transformed Hamiltonian $\hat{H}_{\mathrm{QDM},\mathrm{L}} = \hat{H}_{1\mathrm{L}} + \hat{H}_{2\mathrm{L}}$, where
\begin{gather*}
  \hat{H}_{\beta\mathrm{L}}=\phi(t)e^{-i\omega_Lt}
  V_{1,0}^{(\beta\mathrm{L})}a_\beta^\dagger + \mathrm{H.c.}
,\quad
\end{gather*}
index $\beta$ enumerates the states of QDM $|1\rangle\rangle$ and $|2\rangle\rangle$,
and the new matrix elements are related to the old ones though the transformation matrix $S_\vartheta$ defined in Eq.~\eqref{e:transf} \emph{via}
\begin{equation*}
    \Bigg(
      \begin{array}{c}
        V_{1,0}^{(1\mathrm{L})} \\
        V_{1,0}^{(2\mathrm{L})}
      \end{array}
    \Bigg) = S_\vartheta
    \Bigg(
      \begin{array}{c}
        V_{1\mathrm{I},0\mathrm{I}}^{(\mathrm{L})} \\
        V_{1\mathrm{II},0\mathrm{II}}^{(\mathrm{L})}
      \end{array}
    \Bigg).
\end{equation*}
Here, the interband matrix element is given by $V_{1\alpha,0\alpha}^{(\eta)} = -e\sqrt{2}\big|\mathbf r_{cv}^{(\alpha)}\big|$.

The transformed Hamiltonian of the QDM interaction with the emitted photons of frequencies $\omega_{\mathrm{1R}}$ and $\omega_{\mathrm{2R}}$ is also a sum of two terms, $\hat{H}_{\mathrm{QDM},\mathrm{R}}=\hat{H}_{1\mathrm{R}}+\hat{H}_{2\mathrm{R}}$, with
\begin{equation*}
    \hat{H}_{\beta\mathrm{R}} = i B_\beta^{\vphantom{\dagger}}a_\beta^\dagger
    -i B_\beta^\dagger a_\beta^{\vphantom{\dagger}}
\end{equation*}
and the new operators are given by
\begin{equation*}
    \Bigg(
      \begin{array}{c}
        B_1 \\
        B_2
      \end{array}
    \Bigg) = S_\vartheta
    \Bigg(
      \begin{array}{c}
        g_{\mathrm{I},1\mathrm{R}}^{\vphantom{+}} V_{\mathrm{1I,0I}}^{(1\mathrm{R})}b_{1\mathrm{R}}^{\vphantom{+}}\\
        g_{\mathrm{II},2\mathrm{R}}^{\vphantom{+}} V_{\mathrm{1II,0II}}^{(2\mathrm{R})}b_{2\mathrm{R}}^{\vphantom{+}}
      \end{array}
    \Bigg).
\end{equation*}

\section{Photoluminescence from a quantum-dot molecule}

\begin{figure}[!ht]
\includegraphics[width=6cm]{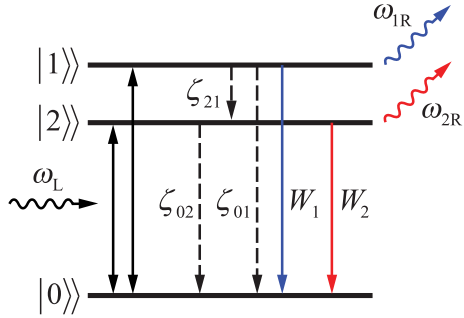}
\caption{(Color online) Energy-level diagram and transitions in a QDM comprising a pair of coherently coupled QDs.
The classical optical field of frequency $\omega_\mathrm{L}$ excites the low-energy state $|2\rangle\rangle$ and/or the high-energy state $|1\rangle\rangle$ of the QDM.
The excited states then decay at rates $\zeta_{01}$ and $\zeta_{02}$ nonradiatively, or at rates $W_1$ and $W_2$ with the emission of secondary photons $\omega_{1\mathrm{R}}$ and $\omega_{2\mathrm{R}}$.
Solid and dashed arrows correspond to the radiative and nonradiative transitions, respectively; $\zeta_{21}$ is the rate of transitions $|1\rangle\rangle\to|2\rangle\rangle$.}\label{f:scheme}
\end{figure}

We next use the results of the previous section to calculate the intensity of the photoluminescence from the QDM comprising QDs with resonant electronic subsystems.
The energy-level diagram illustrating the excitation of the QDM and its radiative and nonradiative relaxation channels is shown in Fig.~\ref{f:scheme}.
Both the interband and intraband nonradiative relaxations, shown by the dashed arrows, occur due to the interaction of the QDM with a bath and are described by the phenomenological rates $\zeta_{01}$, $\zeta_{02}$, and $\zeta_{21}$.
In order to calculate the rates $W_1$ and $W_2$ of the spontaneous light emission from states $|1\rangle\rangle$ and $|2\rangle\rangle$, we construct a five-by-five density matrix using the following basis:
\begin{equation*}
  |1)=|00\rangle|\mathrm{0R}\rangle,\quad
  |2)=|1\rangle\rangle|\mathrm{0R}\rangle,\quad
  |3)=|2\rangle\rangle|\mathrm{0R}\rangle,\quad
  |4)=|00\rangle|\mathrm{1R}\rangle,\quad
  |5)=|00\rangle|\mathrm{2R}\rangle,
\end{equation*}
where $|\mathrm{0R}\rangle$ denotes the vacuum of photons and $|\mathrm{1R}\rangle$ and $|\mathrm{2R}\rangle$ are the states of the emitted photons.

The dynamics of the QDM is governed by the generalized master equation for the reduced density matrix\cite{Blum_1981}
\begin{equation}\label{e:kin}
   \frac{\partial\rho_{ij}}{\partial t}=-\,\frac{i}{\hbar} \big[\hat{H},\hat{\rho}\big]_{ij}
   +\delta_{ij}\sum_{k\neq j}\zeta_{jk}\rho_{kk}
   -\gamma_{ij}\rho_{ij},
\end{equation}
where $\delta_{ij}$ is the Kronecker delta, $\zeta_{jk}$ denotes the rate of transitions $|k\rangle\rangle\to|j\rangle\rangle$ due to the thermal interaction with the bath, $\gamma_{ij}=(\gamma_{ii} + \gamma_{jj})/2 + \bar{\gamma}_{ij}$ for $i\neq j$ gives the damping rate of the $\rho_{ij}$ coherence, $\gamma_{ii}$ is the total decay rate of population out of state $|i)$, and $\bar{\gamma}_{ij}$ is the pure dephasing rate of transition $|j)\to|i)$.
We assume that the pure dephasing rate is the same for all transitions $\bar{\gamma}_{ij} \equiv \bar{\gamma}$, and its temperature dependence is given by the phenomenological formula\cite{Kruchinin_2010_PRB_81_245303,Kruchinin_2008_PRB_78_125311}
\begin{equation}\label{e:deph}
  \bar\gamma(T)=\gamma_0+aT
  +b\left[\exp\left(\frac{\hbar\omega_{\mathrm{LO}}}{k_\mathrm{B}T}\right)-1\right]^{-1},
\end{equation}
where $\gamma_0$ is the dephasing rate due to the radiative and nonradiative transitions induced by the interaction with the bath and $\hbar\omega_{\mathrm{LO}}$ is the energy of the longitudinal optical (LO) phonons in QDs.
The last two terms in this expression describe the interaction of the QDM with the acoustic and LO phonons through the phenomenological coefficients $a$ and $b$.
Since the typical energy splitting of the QDM states due to the interdot Coulomb interaction is of the order of the cutoff energy (a few millielectronvolts) of the acoustic phonon dispersion~\cite{Fedorov_2002_SSC_122_139}, and much smaller than the LO phonon energy (tens of millielectronvolts), the upper state $|1\rangle\rangle$ nonradiatively decays to the lower state $|2\rangle\rangle$ predominantly with the emission of acoustic phonons.

By considering the stationary excitation  $(\phi=\mathrm{const})$ and perturbatively solving Eq.~\eqref{e:kin} to the lowest orders in the electron--photon interaction, one can find the photon emission rates $W_1=\partial\rho_{44}/\partial t$ and $W_2=\partial\rho_{55}/\partial t$.
The measurable luminescence differential cross section (LDCS), which gives the energy emitted by the QDM per a unit solid angle $\mathrm{d}\Omega$ in a unit frequency interval $\mathrm{d}\omega_{i\mathrm{R}}$, scales in proportion to the photon emission rate and is given by
\begin{equation}
    \frac{\mathrm{d}\sigma_i}{\mathrm{d}\Omega\mathrm{d}\omega_{i\mathrm{R}}}=
    \frac{V\hbar\omega^3_{i\mathrm{R}}}{4(\pi c)^3}\frac{W_i}{I_\mathrm{L}}\quad(i=1\,\,\mathrm{or}\,\,2),
\end{equation}
where $I_\mathrm{L}$ is the excitation light intensity.
With this relationship, the major contributions to the LDCS from the excited states of the QDM are found to be
\begin{subequations}\label{e:sigma}
\begin{multline}\label{e:sigma1}
  \frac{\mathrm{d}\sigma_1}{\mathrm{d}\Omega \mathrm{d}\omega_{\mathrm{1R}}}\approx
  C(\omega_{\mathrm{1R}})
  \big|V_{\mathrm{1I,0I}}^{\mathrm{(1R)}}\big|^2
  \left[
    \cos^2\vartheta
    \big|V_{1,0}^{\mathrm{(1L)}}\big|^2
    \frac{2}{\gamma_{11}}
    \frac{\gamma_{01}}{\gamma_{01}^2+\Delta_{\mathrm{1,1R}}^2}
    \frac{\gamma_{01}}{\gamma_{01}^2+\Delta_{\mathrm{1,L}}^2}
  \right.\\
  +\left.
    \sin^2\vartheta\,
    \frac{\gamma_{02}}{\gamma_{02}^2+\Delta_{\mathrm{2,1R}}^2}
    \frac{2}{\gamma_{22}}
    \left(
      \big|V_{1,0}^{\mathrm{(2L)}}\big|^2
      \frac{\gamma_{02}}{\gamma_{02}^2+\Delta_{\mathrm{2,L}}^2}
      +\big|V_{1,0}^{\mathrm{(1L)}}\big|^2
      \frac{\zeta_{21}}{\gamma_{11}} \frac{\gamma_{01}}{\gamma_{01}^2+\Delta_{\mathrm{1,L}}^2}
    \right)
  \right]
\end{multline}
and
\begin{multline}\label{e:sigma2}
  \frac{\mathrm{d}\sigma_2}{\mathrm{d}\Omega \mathrm{d}\omega_{\mathrm{2R}}}\approx
  C(\omega_{\mathrm{2R}})
  \big|V_{\mathrm{1II,0II}}^{\mathrm{(2R)}}\big|^2
  \left[
    \sin^2\vartheta
    \big|V_{1,0}^{\mathrm{(1L)}}\big|^2
    \frac{2}{\gamma_{11}}
    \frac{\gamma_{01}}{\gamma_{01}^2+\Delta_{\mathrm{1,2R}}^2}
    \frac{\gamma_{01}}{\gamma_{01}^2+\Delta_{\mathrm{1,L}}^2}
  \right.\\
  +\left.
    \cos^2\vartheta\,
    \frac{\gamma_{02}}{\gamma_{02}^2+\Delta_{\mathrm{2,2R}}^2}
    \frac{2}{\gamma_{22}}
    \left(
      \big|V_{1,0}^{\mathrm{(2L)}}\big|^2
      \frac{\gamma_{02}}{\gamma_{02}^2+\Delta_{\mathrm{2,L}}^2}
      +\big|V_{1,0}^{\mathrm{(1L)}}\big|^2 \frac{\zeta_{21}}{\gamma_{11}}
      \frac{\gamma_{01}}{\gamma_{01}^2+\Delta_{\mathrm{1,L}}^2}
    \right)
  \right],
\end{multline}
\end{subequations}
where $C(\omega)=4\omega^4/(\pi c^4\hbar^2)$ and $\Delta_{i,j}=\omega_i-\omega_j$.
The intensity of the QDM photoluminescence is the sum of these two contributions.

\section{Numerical results}

We illustrate the results obtained by considering two identical, 4-nm in diameter QDs made from the cubic modification of CdSe, which is characterized by the following set of material parameters:~\cite{Norris_1996_PRB_53_16338}
$m_c^{(\alpha)} = 0.11~m_0$ ($m_0$ is the free-electron mass),
$m_v^{(\alpha)}=1.14~m_0$,
$E_g^{(\alpha)}=1736$~meV,
$P_\alpha=1.48\times10^{-19}$~cm$^3$~g~s$^{-2}$,
$\hbar\omega_\mathrm{LO}=26$~meV,
and $\varepsilon_\alpha=5.8$ for $\alpha = \mathrm{I}$ or II.
The QDM is assumed to be embedded in fused silica with $\varepsilon_\mathrm{M} = 2.13$ and the relaxation parameters are chosen to be $\gamma_0=7.7\times10^7$~s$^{-1}$, $a=1.5\times10^{10}$~s$^{-1}$K$^{-1}$, $b=2.3\times10^{10}$~s$^{-1}$, and $\gamma_{01}=\gamma_{02}=40~\mu$eV.
According to Eqs.~\eqref{e:M}, \eqref{e:chi}, \eqref{e:03}, and \eqref{e:sigma}, the photoluminescence intensity heavily depends on the mutual orientations of the transition dipole moments and the polarization of the excitation field.
For the sake of definiteness, we focus on the ideal situation in which the three vectors $\mathbf r_\mathrm{I}$, $\mathbf r_\mathrm{II}$, and $\mathbf e_\mathrm{L}$ are codirectional, and thus, both the interdot and the QDM--light interactions are strongest.

\begin{figure}[!ht]
\includegraphics[width=9cm]{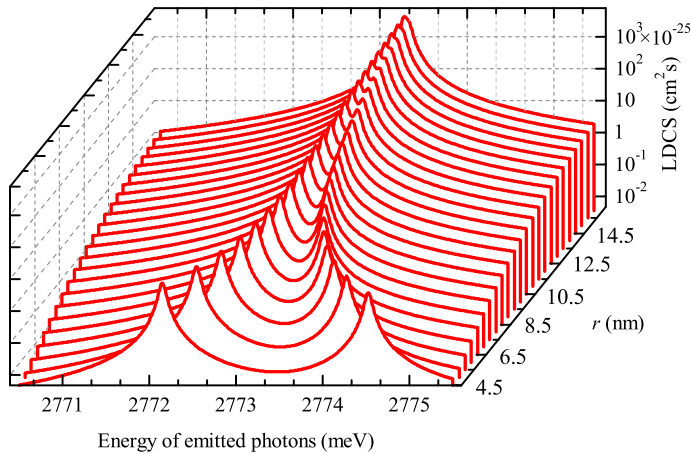}
\caption{(Color online) Photoluminescence spectrum of a QDM for different distances $r$ between the QD centers [the spectrum is the sum of LDCSs given in Eqs.~\eqref{e:sigma1} and \eqref{e:sigma2}].
The QDs are assumed to be identical, made of CdSe, and have radii of 2~nm.
The relaxation rates are $\bar\gamma(4~\mathrm{K})=6\times10^{10}$~s$^{-1}$, $\zeta_{01}=\zeta_{02}=\zeta_{12}/2=10^8$~s$^{-1}$, and $\gamma_{01}=\gamma_{02}=40~\mu$eV.
For other parameters refer to the text.
}\label{f:LDCS_r}
\end{figure}

Figure~\ref{f:LDCS_r} shows how the photoluminescence spectrum of the QDM changes with the interdot distance when the excitation energy $\hbar\omega_\mathrm{L}=2773.34$~meV coincides with the fundamental transition energy in the decoupled QDs.
The two peaks, centered at frequencies $\omega_1$ and $\omega_2$, arise in the spectrum due to the coherent coupling between the QDs.
The peaks are seen to grow with $r$ as they gradually merge together and approach the excitation frequency.
The peaks' splitting scales as $1/r^3$ [see Eqs.~\eqref{e:M} and \eqref{e:03}] whereas the full width at half maximum (FWHM) of both peaks is about $2\gamma_{01} = 80~\mu$eV [see Eq.~\eqref{e:sigma} with $\vartheta=\pi/4$].
The splitting is seen to exceed 2~meV when the QDs nearly touch each other, and becomes too small for experimental resolution when $r$ exceeds 10~nm.
By carefully choosing the parameters of the QDs, interdot distance, and excitation frequency, one can tune the positions and relative intensities of the QDM photoluminescence peaks as desired for practical applications.

\section{Conclusions}

We have developed a theory of low-temperature, stationary photoluminescence from a pair of spherical quantum dots coupled  by the Coulomb interaction in a quantum-dot molecule.
The lowest-energy electron--hole-pair states of the dots were assumed to be nearly resonant and characterized by low decay and dephasing rates.
The coherent coupling of the quantum dots under these conditions was shown to manifest itself in the molecule's photoluminescence spectrum as a pair of peaks, the intensities and spectral positions of which are determined by the geometry and material of the nanocrystals, as well as by the rates of the energy and phase relaxations of their electronic subsystems.
We also derived an expression for the photoluminescence differential cross section, which is useful for interpreting and analyzing the secondary emission spectra of coherently coupled quantum nanostructures.

\section*{Acknowledgements}

The authors gratefully acknowledge the financial support of this work from the Ministry of Education and Science of the Russian Federation (Grant No.~14.B25.31.0002) and the Russian Foundation for Basic Research (Grants No.~12-02-01263 and No.~12-02-00938).
The Ministry of Education and Science of the Russian Federation also supports A.S.B. and M.Yu.L., through its scholarships of the President of the Russian Federation for young scientists and graduate students (2013--2015).
A.S.B. is also grateful to the Dynasty Foundation Support Program for Physicists.
The work of I.D.R. is sponsored by the Australian Research Council, through its Discovery Early Career Researcher Award DE120100055.

\bibliographystyle{apsrev4-1}
\bibliography{article}

\end{document}